\documentclass[conference]{IEEEtran}
\IEEEoverridecommandlockouts
\usepackage{cite}
\usepackage{amssymb,amsfonts}
\usepackage{amsmath}
\usepackage{algorithmic}
\usepackage{graphicx}
\usepackage{textcomp}
\usepackage{xcolor}
\usepackage{empheq}
\usepackage{subfig}
\usepackage{algorithm}
\usepackage{bm}
\usepackage{hyperref}
\usepackage[font=footnotesize]{caption}  

\def\BibTeX{{\rm B\kern-.05em{\sc i\kern-.025em b}\kern-.08em
    T\kern-.1667em\lower.7ex\hbox{E}\kern-.125emX}}
\begin{document}

\title{Predictive Target-to-User Association in Complex Scenarios via Hybrid-Field ISAC Signaling 
}

\author{\IEEEauthorblockN{Yifeng Yuan\IEEEauthorrefmark{1}, Miaowen Wen\IEEEauthorrefmark{2}, Xinhu Zheng\IEEEauthorrefmark{1}, Shuoyao Wang\IEEEauthorrefmark{3}, and Shijian Gao\IEEEauthorrefmark{1}}
\IEEEauthorblockA{
\IEEEauthorrefmark{1}IoT Thrust, The Hong Kong University of Science and Technology (Guangzhou), China
}
\IEEEauthorblockA{\IEEEauthorrefmark{2}South China University of Technology, Guangzhou, China 
}
\IEEEauthorblockA{\IEEEauthorrefmark{3}College of Electronic and Information Engineering, Shenzhen University, China 
}
}
\maketitle

\begin{abstract}
This paper presents a novel and robust target-to-user (T2U) association framework to support reliable vehicle-to-infrastructure (V2I) networks that potentially operate within the hybrid field (near-field and far-field). To address the challenges posed by complex vehicle maneuvers and user association ambiguity, an interacting multiple-model filtering scheme is developed, which combines coordinated turn and constant velocity models for predictive beamforming. Building upon this foundation, a lightweight association scheme leverages user-specific integrated sensing and communication (ISAC) signaling while employing probabilistic data association to manage clutter measurements in dense traffic. Numerical results validate that the proposed framework significantly outperforms conventional methods in terms of both tracking accuracy and association reliability.
\end{abstract}

\section{Introduction}
Integrated sensing and communications (ISAC) has emerged as a promising technology for future 6G networks, enabling simultaneous communication and sensing capabilities through shared hardware architectures and spectrum resources \cite{liu2022integrated}. This integration significantly improves spectrum efficiency and reduces system complexity \cite{cheng2022integrated,nguyen2023multiuser}. In vehicle-to-everything (V2X) scenarios, ISAC simultaneously addresses critical demands for reliable high-speed connectivity and precise environmental sensing \cite{duan2020emerging,noor20226g}. Such dual functionality is particularly essential for vehicle-to-infrastructure (V2I) communications, where roadside units (RSUs) must maintain reliable connections with multiple vehicles in dynamic environments.

In V2I communications, beam alignment plays a vital role in securing reliable high-speed links between RSUs and vehicles. By leveraging ISAC capabilities\cite{liu2020radar,yuan2020bayesian}, RSUs can extract vehicle state information directly from radar echoes to assist beam tracking, which significantly reduces communication overhead compared to conventional pilot-based approaches\cite{va2016beam}. However, in multi-user scenarios with complex road geometries and diverse vehicle maneuvers, a fundamental challenge arises: the target-to-user (T2U) association problem. Additionally, due to the large-scale antenna arrays operating at mmWave frequencies, vehicles tend to traverse across either the near-field or the far-field regions of RSUs, necessitating the beamforming in hybrid field. \cite{li2023sensing}. 

Despite extensive research on ISAC-based T2U association, existing approaches encounter significant limitations in practical scenarios. Previous studies \cite{liu2020radar,yuan2020bayesian} assumed steady vehicle movement along straight roads parallel to the RSU array, oversimplifying real-world road geometries and vehicle maneuvers. Although recent research has attempted to address this limitation by incorporating vehicle motion information through uplink pilots \cite{guo2024predictive} or exploiting road prior knowledge \cite{meng2023vehicular}, these solutions either incur additional overhead or require complex coordinate transformations.  More critically, existing association methods demonstrate significant shortcomings in handling realistic multi-user scenarios. Current approaches primarily adopt distance metrics like  Euclidean distance \cite{liu2020radar} or Kullback-Leibler divergence (KLD) \cite{wang2022multi} to characterize the correlation between radar measurements and communication users. However, these methods overlook the presence of non-cooperative vehicles that share similar motion patterns with users. From a radar sensing perspective, echo signals from these non-cooperative vehicles typically act as clutters, which will significantly complicate the T2U association process \cite{chen2023multiuser}. To mitigate this issue, existing solutions rely on external units, such as the reconfigurable intelligent surface (RIS) mounted on vehicles or visual sensor installed at the RSU \cite{mizmizi2023target,zhang2024integrated,cazzella2024deep}, at the cost of hardware expenditure.

In this paper, a robust T2U association framework is proposed for hybrid-field multi-vehicle ISAC systems that effectively addresses the aforementioned challenges. Our contributions are twofold. First, we develop an interacting multiple-model (IMM) based predictive tracking scheme \cite{blom1988interacting}, which incorporates coordinated turn (CT) model to accurately capture complex vehicle maneuvers and enable reliable state estimation under diverse road geometries. Second, we propose a novel ISAC-based association framework that leverages both the user-specific communication information embedded in radar echoes and probabilistic data association techniques to handle dense traffic scenarios with multiple clutter vehicles \cite{bar2009probabilistic}. Through numerical experiments, the superiority of the proposed framework is validated in terms of both tracking accuracy and association reliability in realistic traffic scenarios. This makes the devised association scheme an appealing solution for supporting high-speed V2I connection.

\begin{figure}[tbp]
\centerline{\includegraphics[width=2.5in]{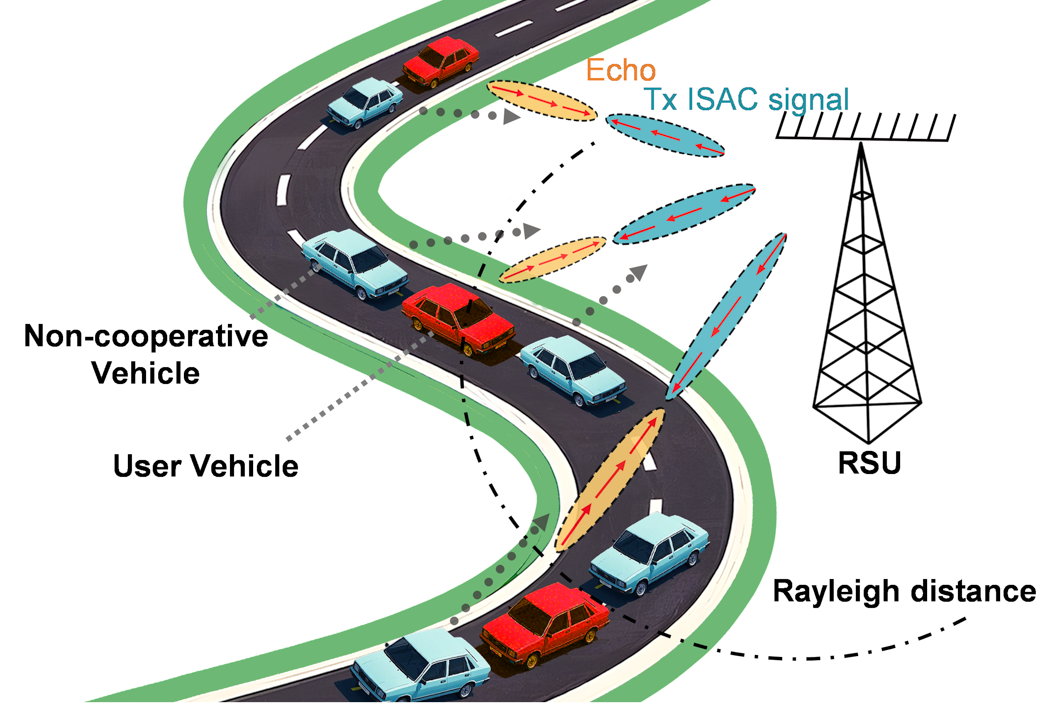}}
\caption{ Illustration of studied hybrid-field ISAC V2I scenario.}
\label{fig_scenario}
\vspace{-6pt}
\end{figure}

\section{System Model}
As illustrated in Fig. \ref{fig_scenario}, a hybrid-field ISAC system is considered, where an RSU serves $K$ communication vehicles. In the vicinity of each served vehicle, there are $M_k$ non-cooperative vehicles whose radar echoes are also captured by the RSU. The RSU employs a uniform linear array (ULA) with $N_t$ transmit and $N_r$ receive antennas, spaced at distance $d$. Under an array aperture of $D = (N_t-1)d$, the Rayleigh distance $\frac{2D^2}{\lambda}$ demarcates the boundary of the near-field and the far-field regions, with $\lambda$ being the carrier wavelength.
\vspace{-1mm}
\subsection{Channel Model}
Given that vehicular users operate in both near-field and far-field regions, we adopt a unified spherical wave model to characterize the RSU-object propagation in our mmWave massive antenna system, assuming line-of-sight (LoS) dominance.
For a ULA with its $n$-th element positioned at $\mathbf{s}_n = [nd, 0]^T$, $n \in \{0,\ldots,N_t-1\}$, we consider an object at position $\mathbf{p} = [r\cos\theta, r\sin\theta]^T$, where $r$ and $\theta$ denote the distance and angle relative to the array center, respectively. The distance between the $n$-th antenna element and the object is given by:
\begin{equation}
r_{n}(r,\theta) = \Vert \mathbf{p}-\mathbf{s}_n \Vert_2 = \sqrt{r^2 + n^2d^2 - 2rnd\cos\theta}.
\end{equation}
Under the assumption of identical channel gains between antenna elements and the user \cite{wang2023near}, the channel vector $\mathbf{h} \in \mathbb{C}^{N_t \times 1}$ between the RSU and the object can be expressed as
\begin{equation}
\setlength{\abovedisplayskip}{3pt}
\setlength{\belowdisplayskip}{3pt}
\mathbf{h} = \alpha\mathbf{a}(r,\theta),
\end{equation}
where $\alpha = \sqrt{\frac{\lambda}{4\pi r}}e^{-j\frac{2\pi}{\lambda}r}$ represents the complex channel gain, and $\mathbf{a}(r,\theta)$ denotes the array steering vector with entries $[\mathbf{a}(r,\theta)]_n=e^{-j\frac{2\pi}{\lambda}(r_n(r,\theta)-r)}$. The unified model ensures accurate channel characterization across the above hybrid field.

\vspace{-1mm}
\subsection{ISAC Signaling}
\subsubsection{Communication Signal}
At each tracking epoch, the RSU transmits multi-beam ISAC signals to serve $K$ users. The transmitted signal at the $l$-th epoch can be expressed as
\vspace{-1mm}
\begin{equation}
    \tilde{\mathbf{s}}_l\left(t\right) = \sqrt{p_l} \mathbf{F}_l\mathbf{s}_l\left(t\right)\in \mathbb{C}^{N_t\times1},
    \vspace{-1mm}
\end{equation}
where $\mathbf{s}_l(t) = [s_{1,l}(t),...,s_{K,l}(t)]^T$ denotes the $K$ downlink ISAC streams, $\mathbf{F}_l\in\mathbb{C}^{N_t\times K}$ represents the transmit beamforming matrix, with its $k$-th column $\mathbf{f}_{k,l}$ representing the beamforming vector for the $k$-th user, and $p_l$ is the transmit power at the RSU.
Given the narrow beamwidth of the large-scale antenna array, inter-beam interference can be neglected \cite{liu2020radar}. Thus, the received signal at the $k$-th user is
\vspace{-1mm}
\begin{equation}
r_{k,l}(t) \!=\! \sqrt{p_l} \alpha_{k,l} \mathbf{a}^H\!(r_{k,l},\theta_{k,l})\mathbf{f}_{k,l}s_{k,l}\!\left(t\right)e^{j2\pi\varrho_{k,l}t}\!+\!z_c(t),
\vspace{-1mm}
\end{equation}
where $\varrho_{k,l}$ denote the Doppler frequency and $z_c(t)$ represents the zero-mean complex Gaussian noise with variance $\sigma_c^2$.
Assuming unit power ISAC streams, the receive signal-to-noise ratio (SNR) for the $k$-th user is given by:
\vspace{-1mm}
\begin{equation}
\text{SNR}_{k,l} = \frac{p_l|\alpha_{k,l}\mathbf{a}^H(r_{k,l},\theta_{k,l})\mathbf{f}_{k,l}|^2}{\sigma_c^2},
\vspace{-1mm}
\end{equation}
so the achievable sum-rate of all $K$ users is
\begin{equation}
\setlength{\abovedisplayskip}{3pt}
\setlength{\belowdisplayskip}{3pt}
R_l = \sum_{k=1}^K \log_2(1+\text{SNR}_{k,l}).
\end{equation}
\subsubsection{Echo Signal}
For the $k$-th user's beam coverage, which includes both the intended user and $M_k$ clutter vehicles, the echo signal after received beamforming can be expressed as
\begin{equation}
\setlength{\abovedisplayskip}{3pt}
\setlength{\belowdisplayskip}{3pt}
\begin{aligned}
r_{k,l}(t) = & \sum_{m=0}^{M_k} \beta_{k,l}^{(m)}\mathbf{w}_{k,l}^H\mathbf{b}(r_{k,l}^{(m)},\theta_{k,l}^{(m)})\mathbf{a}^H(r_{k,l}^{(m)},\theta_{k,l}^{(m)})\mathbf{f}_{k,l} \\
& \times s_{k,l}(t-\tau_{k,l}^{(m)})e^{j2\pi\mu_{k,l}^{(m)}t} + z_r(t),
\end{aligned}
\end{equation}
where $\mathbf{w}_{k,l}$ is the receive beamforming vector, $\mathbf{b}(r_{k,l}^{(m)},\theta_{k,l}^{(m)})$ is the receiving steering vector, and $z_r(t)$ denotes the complex Gaussian noise with variance $\sigma_r^2$. The echo sources comprise both the intended user ($m=0$) and clutter vehicles ($m=1,\ldots,M_k$), characterized by their reflection coefficients $\beta_{k,l}^{(m)}$, locations $({r_{k,l}^{(m)},\theta_{k,l}^{(m)}})$, propagation delays $\tau_{k,l}^{(m)}$, and Doppler shifts $\mu_{k,l}^{(m)} = \frac{2v^{(m)}\cos(\theta_{k,l}^{(m)} - h_{k,l}^{(m)})}{\lambda}$, where $v^{(m)}$ and $h_{k,l}^{(m)}$ represent the velocity magnitude and the heading angle of the $m$-th target\cite{gao2020estimating}, respectively. 
\vspace{-1mm}

\begin{figure*}[tbp]
\centering
\subfloat[]{\includegraphics[width=4.5cm]{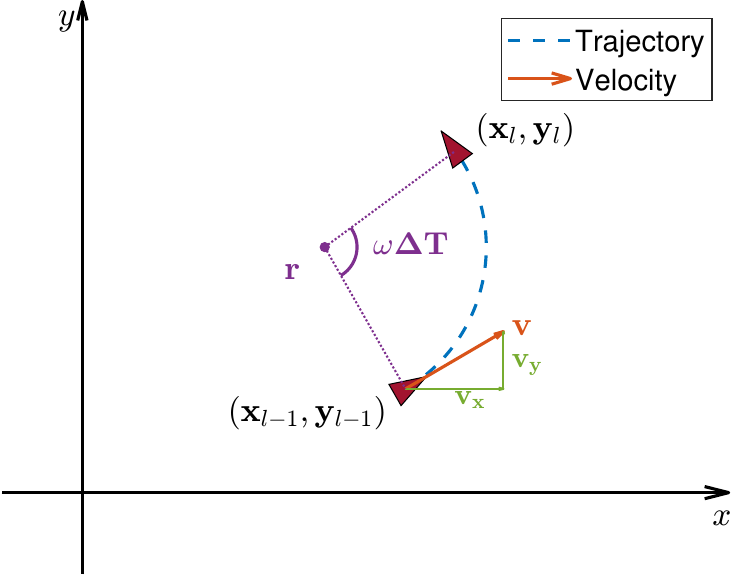}} 
\hfil
\subfloat[]{\includegraphics[width=9cm]{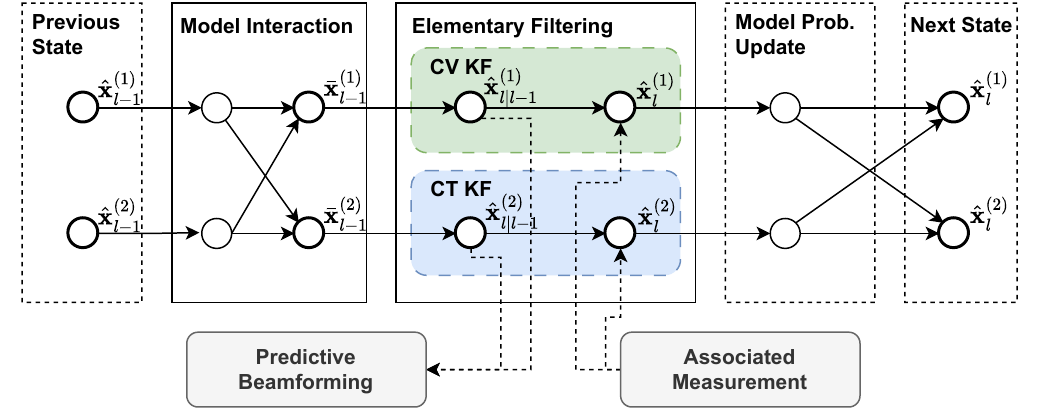}} 
\hfil
\subfloat[]{\includegraphics[width=2.75cm]{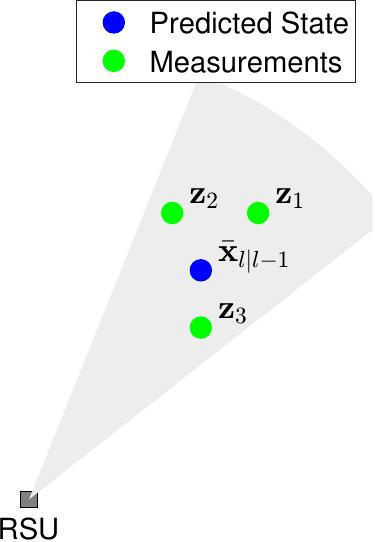}} 
\caption{ Illustration of the proposed predictive T2U association framework: (a) CT motion model, (b) Process of the IMM filtering, (c) Potential cluttered measurements within a beam coverage.}
\label{fig_ct_imm}
\vspace{-5mm}
\end{figure*}

\section{Predictive Beamforming via IMM Filter}
In this section, we develop a predictive beamforming framework to tackle the challenge arising from the complex vehicle motion, which serves as the foundation for the proposed target-to-user association strategy.
\vspace{-1mm}
\subsection{Coordinated Turn State Evolution Model}
The state evolution model directly influences the effectiveness of vehicle state tracking. In practical complex scenarios, vehicles often perform maneuvers and turning, which prompt us to introduce the CT model to characterize vehicle motion \cite{roth2014ekf}. The CT model assumes that targets move in a plane with a constant angular velocity, thereby offering a more realistic representation of vehicle turning behavior. 

The state vector per user at the $l$-th epoch is defined as 
\vspace{-1mm}
\begin{equation}
    \mathbf{x}_l = [x_l, y_l, v_{x,l}, v_{y,l}, \omega_l]^T,
    \vspace{-1mm}
\end{equation}
where $(x_l, y_l)$ denotes the vehicle position in Cartesian coordinates, $(v_{x,l}, v_{y,l})$ represents the velocity components, and $\omega_l$ is the turn rate that characterizes the vehicle's rotational motion. Based on this state representation, the vehicle's heading angle can be derived as $h_l = \text{atan2}(v_{y,l}, v_{x,l})$, which evolves according to the turning rate $\frac{d{h_l}}{dt} = \omega_l$.

As illustrated in Fig. \ref{fig_ct_imm}(a), we model the vehicle's movement as circular motion with constant turn rate within each tracking epoch. Specifically, both the velocity magnitude $v_l = \sqrt{v_{x,l}^2 + v_{y,l}^2}$ and the turn rate $\omega_l$ are assumed to remain constant during the epoch, resulting in circular motion with a radius $r_l = v_l/\omega_l$. Under these assumptions, the state evolution for an epoch duration of $\Delta T$ can be characterized by:
\begin{equation}
\setlength{\abovedisplayskip}{4pt}
\setlength{\belowdisplayskip}{4pt}
\hspace{-2mm}
\begin{cases}
\begin{aligned}
&x_{l+1}\!=\!x_l \!+\! \frac{\sin(\omega_l\Delta \!T)}{\omega_l}v_{x,l} \!+\! \frac{\cos(\omega_l\Delta \!T)\!-\!1}{\omega_l}v_{y,l} \!+ \!w_x, \\
&y_{l+1} \!= \!y_l \!+\! \frac{1\!-\!\cos(\omega_l\Delta \!T)}{\omega_l}v_{x,l} \!+\! \frac{\sin(\omega_l\Delta\! T)}{\omega_l}v_{y,l} \!+ \!w_y, \\
&v_{x,l+1} = v_{x,l}\cos(\omega_l\Delta T) - v_{y,l}\sin(\omega_l\Delta T) + w_{v_x}, \\
&v_{y,l+1} = v_{x,l}\sin(\omega_l\Delta T) + v_{y,l}\cos(\omega_l\Delta T) + w_{v_y}, \\
&\omega_{l+1} = \omega_l + w_{\omega},   \label{eqn:state_model}
\end{aligned}
\end{cases}
\end{equation}
where the system process noise terms $[w_x, w_y, w_{v_x}, w_{v_y}, w_{\omega}]^T$ characterize the uncertainties arising from vehicle acceleration and model imperfections. These noise components are driven by random accelerations in the $x$-, $y$-, and angular-directions, all modeled as zero-mean Gaussian variables. 
\vspace{-1mm}
\subsection{IMM Filtering Framework}
\vspace{-1mm}
Inspired by the IMM-based tracking approach in complex road conditions \cite{meng2023vehicular}, we utilize an IMM filtering framework that incorporates multiple motion models simultaneously to track vehicle states. We employ two complementary motion models: the CT model for capturing turning and lane-changing maneuvers, and the constant velocity (CV) model which is a special case obtained by setting $\omega = 0$, for tracking straight-path motion with constant velocity.
The IMM filter recursively estimates vehicle states by leveraging both current measurements and prior knowledge. At each epoch $l$, two motion models are integrated to generate state estimates $\hat{\mathbf{x}}_l$ from previous estimates $\hat{\mathbf{x}}_{l-1}$.
The IMM filter operates through three key steps elaborated as follows:
\subsubsection{Model Interaction}
To manage the uncertainty in vehicle motion patterns, the state estimates and covariances from different models need to be optimally combined. The mixing weights for this combination are derived from two key probabilities: the \textit{model probability} $\rho^{(j)}_{l-1}$, which represents the likelihood of each model being correct at the previous epoch, and the \textit{transition probability} $\pi_{j,i}$, indicating the likelihood of switching from model $j$ to model $i$. The mixing weights are calculated as follows:
\begin{equation}
\setlength{\abovedisplayskip}{3pt}
\setlength{\belowdisplayskip}{3pt}
    c^{(i|j)}_l = \frac{\pi_{j,i}\rho^{(j)}_{l-1}}{\sum_{j=1}^2 \pi_{j,i}\rho^{(j)}_{l-1}}.
\end{equation}
Based on these weights, the mixed inputs for each model are computed as
\begin{equation}
\begin{aligned}
\setlength{\abovedisplayskip}{3pt}
\setlength{\belowdisplayskip}{3pt}
&\bar{\mathbf{x}}^{(i)}_{l-1} = \sum_{j=1}^2 \hat{\mathbf{x}}^{(j)}_{l-1}c^{(i|j)}_l, \\
&\bar{\mathbf{M}}^{(i)}_{l-1} \!=\! \sum_{j=1}^2 \!c^{(i|j)}_l\![\mathbf{M}^{(j)}_{l-1} \!+\! (\bar{\mathbf{x}}^{(i)}_{l-1} \!-\! \hat{\mathbf{x}}^{(j)}_{l-1})(\bar{\mathbf{x}}^{(i)}_{l-1} \!-\! \hat{\mathbf{x}}^{(j)}_{l-1})^T], 
\end{aligned}
\end{equation}
where $\bar{\mathbf{x}}^{(i)}_{l-1}$ and $\bar{\mathbf{M}}^{(i)}_{l-1}$ denote the mixed state and covariance corresponding to the $i$-th model filter.
\vspace{-1pt}
\subsubsection{Elementary Filtering}
Each model independently performs Extended Kalman Filtering (EKF) to recursively estimate the state. The process begins with predicting the state using $\hat{\mathbf{x}}^{(i)}_{l|l-1} = \mathbf{g}^{(i)}(\bar{\mathbf{x}}^{(i)}_{l-1})$, where $\mathbf{g}^{(i)}(\cdot)$ is the state evolution model defined in (\ref{eqn:state_model}).
Following the prediction, the state is updated with the new measurement using the equation $\hat{\mathbf{x}}^{(i)}_l = \hat{\mathbf{x}}^{(i)}_{l|l-1} + \mathbf{K}_l(\mathbf{z}_l - \mathbf{h}(\hat{\mathbf{x}}_{l|l-1}))$, where $\mathbf{K}_l$ denotes the Kalman gain, $\mathbf{z}_l$ represents the measurement vector, and $\mathbf{h}(\cdot)$ is the measurement function, which will be specified later.
\vspace{-1pt}
\subsubsection{Model Probability Update}
The model probabilities $\rho^{(i)}_l$ at the $l$-th epoch are updated using the measurement likelihood function $L^{(i)}_l$, which quantifies how well each model $i$ matches the current observation. This function is derived from the measurement residual and the residual covariance. The update equation for the model probabilities is given by:
\begin{equation}
\setlength{\abovedisplayskip}{3pt}
\setlength{\belowdisplayskip}{3pt}
\rho^{(i)}_l = \frac{L^{(i)}_l\sum_{j=1}^2 \pi_{j,i}\rho^{(j)}_{l-1}}{\sum_{i=1}^2 L^{(i)}_l\sum_{j=1}^2 \pi_{j,i}\rho^{(j)}_{l-1}},  \label{eqn:model_prob_update}
\end{equation}
where $\rho^{(j)}_{l-1}$ is the probability of model $j$ at the previous epoch. With these updated model probabilities $\rho^{(i)}_l$ and final state estimates $\hat{\mathbf{x}}^{(i)}_{l|l-1}$, the algorithm can proceed to the next epoch.

\subsection{Predictive Beamforming}
Building on the IMM filtering framework, we develop a predictive beamforming scheme that leverages both CV and CT models for enhanced prediction accuracy. At each tracking epoch, the RSU obtains a combined state prediction as a weighted sum according to the model probabilities. This combined state prediction can be denoted as:
\begin{equation}
\setlength{\abovedisplayskip}{3pt}
\setlength{\belowdisplayskip}{3pt}
   \bar{\mathbf{x}}_{l|l-1} = \sum_{i=1}^2 \bar{\mathbf{x}}^{(i)}_{l|l-1}\rho^{(i)}_{l-1},   \label{eqn:state_predict}
\end{equation}
where $\bar{\mathbf{x}}^{(i)}_{l|l-1}$ represents the state prediction from the $i$-th model and $\rho^{(i)}_{l-1}$ denotes its corresponding probability.

Based on this combined prediction $\bar{\mathbf{x}}_{l|l-1}$, the predicted position can be converted to polar coordinates $\hat{r} = \sqrt{\bar{x}^2 + \bar{y}^2}$ and $\hat{\theta} = \text{atan2}(\bar{y},\bar{x})$. The transmit and receive beamforming vectors for the $k$-th user at the $l$-th epoch are designed as
\begin{subequations}
\setlength{\abovedisplayskip}{3pt}
\setlength{\belowdisplayskip}{3pt}
\begin{align}
\mathbf{f}_{k,l} &= \frac{1}{\sqrt{N_t}}\mathbf{a}\left(\hat{r}_{k,l|l-1},\hat{\theta}_{k,l|l-1}\right) \label{eqn:tx_bf}, \\
\mathbf{w}_{k,l} &= \frac{1}{\sqrt{N_r}}\mathbf{b}\left(\hat{r}_{k,l|l-1},\hat{\theta}_{k,l|l-1}\right), \label{eqn:rx_bf}
\end{align}
\end{subequations}
where $\mathbf{a}(\cdot)$, $\mathbf{b}(\cdot)$ represent the transmit and receive array steering vectors defined in Section II. By leveraging the adaptive capability of the IMM framework, this predictive beamforming approach sets a solid foundation for T2U association. The entire process is illustrated in Fig. \ref{fig_ct_imm}(b).
\vspace{-1mm}
\section{ISAC-Aided Target-to-User Association}
In this section, we propose a probabilistic data association scheme that leverages the advantages of ISAC signaling to effectively address the T2U association problem in multi-vehicle scenarios.
\vspace{-1mm}
\subsection{ISAC-Based Target Detection}
Unlike conventional radar systems that employ a uniform waveform for target detection, ISAC systems transmit user-specific signals. This capability enables separate target detection within each user’s beam coverage. Towards the beam for the $k$-th user, matched filtering is performed by correlating the received signal with the reference waveform $s_{k,l}(t)$ across different Doppler shifts. The matched filter output reaches its maximum when both the delay and Doppler align with the target's parameters. At this peak value, the filtered echo response from the $m$-th target can be expressed as:
\begin{equation}
\setlength{\abovedisplayskip}{3pt}
\setlength{\belowdisplayskip}{3pt}
    \tilde{r}^m_{k,l} = \sqrt{p_l}\sqrt{G}\beta^{(m)}_{k,l}\kappa_{T,l}^{(m)}\kappa_{R,l}^{(m)} + z_r,
\end{equation}
where $G$ denotes the matched-filter gain determined by coherent processing time, and $z_r$ is the output complex white Gaussian noise with variance $\sigma^2_r$. The transmit and receive beamforming gains, denoted as $\kappa_{T,l}^{(m)}$ and $\kappa_{R,l}^{(m)}$ respectively, are defined as follows:
\vspace{-1mm}
\begin{subequations}
    \begin{align}
        \kappa_{T,l} = \mathbf{a}^H(r_{k,l}^{(m)},\theta_{k,l}^{(m)})\mathbf{f}_{k,l}, \\
        \kappa_{R,l} = \mathbf{w}^H_{k,l}\mathbf{b}(r_{k,l}^{(m)},\theta_{k,l}^{(m)}).
        \vspace{-1mm}
    \end{align}
\end{subequations}
For target detection, we employ a constant false alarm rate (CFAR) detector with a predefined false alarm rate $P_{fa}$. Under the noise-only hypothesis $H_0$, the detection threshold is derived as $\eta = -2\sigma_r^2\ln(P_{fa})$. A target is detected when the condition $|\tilde{r}^m_{k,l}|^2 > \eta$ is satisfied.
For each detected target within the $k$-th user's beam coverage, we can extract a measurement vector using established methods such as matched filtering and MUltiple SIgnal Classification (MUSIC) algorithm \cite{meng2023vehicular}, which comprises delay, Doppler, and angle information, giving rise to
\vspace{-1mm}
\begin{equation}
\setlength{\abovedisplayskip}{3pt}
\setlength{\belowdisplayskip}{3pt}
    \mathbf{z}_{k,l}^{(m)} = \begin{bmatrix} \frac{2r_{k,l}^{(m)}}{c}, \mu_{k,l}^{(m)}, \theta_{k,l}^{(m)} \end{bmatrix}^T + \mathbf{z},  \label{eqn:mea_vector}
\end{equation}
where $\mathbf{z} \sim \mathcal{CN}(\mathbf{0},\mathbf{\Sigma})$ represents the measurement noise with covariance matrix $\mathbf{\Sigma} = \text{diag}(\sigma_{\tau}^2, \sigma_{\mu}^2, \sigma_{\theta}^2)$. The variances $\sigma_{\tau}^2$, $\sigma_{\mu}^2$, and $\sigma_{\theta}^2$ are inversely proportional to the output signal-to-noise ratio (SNR) denoted by $\frac{p_lG\beta^2\kappa_{T,l}\kappa_{R,l}}{\sigma^2_r}$.
\vspace{-1mm}
\subsection{Probabilistic Data Association in Cluttered Condition}
As illustrated in Fig. \ref{fig_ct_imm}(c), the detected targets within each beam coverage may originate from either the intended user or non-cooperative vehicles, which creates significant challenges for T2U association. Traditional nearest neighbor approaches struggle in situations where multiple potential targets exhibit similar kinematic behaviors. Motivated by the probabilistic approaches for handling measurement uncertainties in radar systems \cite{bar2009probabilistic},  we apply a probabilistic data association framework that explicitly accounts for measurement origin uncertainties.
Let $\mathcal{Z}_{k,l} = \{\mathbf{z}_{k,l}^{(q)}\}_{q=1}^{Q_{k,l}}$ denote the set of $Q_{k,l}$ measurements detected within the $k$-th user's beam. 
Based on the predicted state $\bar{\mathbf{x}}_{l|l-1}$, we can obtain the predicted measurement $\hat{\mathbf{z}}_{l|l-1} = \mathbf{h}(\bar{\mathbf{x}}_{l|l-1})$. Under the Gaussian assumption of EKF, we can evaluate the likelihood of each measurement originating from the intended user, denoted as 
\vspace{-2mm}
\begin{equation}
    L_l^{(q)} = \frac{N[\mathbf{z}_{k,l}^{(q)};\hat{\mathbf{z}}_{l|l-1},\mathbf{S}_l]}{\lambda},
    \vspace{-2mm}
\end{equation}
where $\mathbf{S}_l = \mathbf{H}\bar{\mathbf{M}}_{l|l-1}\mathbf{H}^T + \mathbf{Q}_m$ represents the measurement residual covariance matrix, $\mathbf{H}$ is the measurement Jacobian matrix linearizing the measurement function, and $\bar{\mathbf{M}}_{l|l-1}$ is the predicted state covariance matrix characterizing the uncertainty in the state prediction.
Assuming the clutter follows a spatial Poisson distribution with density $\lambda$, we can compute the likelihood ratio between target-originated and clutter-originated hypotheses given as $\gamma^{(q)}_l = \frac{L_l^{(q)}}{\lambda}$.
Based on these likelihood ratios, we can compute the association probability for each measurement as
\vspace{-2mm}
\begin{equation}
\beta^{(q)} = \frac{\gamma^{(q)}_l}{\sum_{j=1}^{Q_{k,l}} \gamma^{(j)}_l}, q = 1,\ldots,Q_{k,l}.  \label{eqn:asso_prob}
\vspace{-2mm}
\end{equation}

Finally, we obtain the associated measurement for the $k$-th user through probabilistic combination, denoted as
\begin{equation}
\setlength{\abovedisplayskip}{3pt}
\setlength{\belowdisplayskip}{3pt}
\bar{\mathbf{z}}_{k,l} = \sum_{q=1}^{Q_{k,l}} \beta^{(q)} \mathbf{z}_{k,l}^{(q)}.
\end{equation}
This probabilistic association framework provides a robust solution by explicitly considering the measurement origin uncertainties, enabling reliable T2U association even in dense traffic scenarios with multiple clutter vehicles. This capability enhances the overall user-tracking and communication performance. The overall T2U Association scheme is summarized in Algorithm \ref{alg:association} for reference.

\begin{algorithm}[t]
\caption{Predictive T2U Association Framework}
\label{alg:association}
\begin{algorithmic}[1]
\STATE \textbf{Input: }Model transition probabilities $\{{\pi_{j,i}}\}_{j,i}^2$, measurement noise covariance $\mathbf{Q}_m$, and system process noise $\mathbf{Q}_s$.

\STATE \textbf{Output: }State estimates $\{\hat{\mathbf{x}}_{k,l}^{(i)}\}_{k=1}^K$, beamforming vectors $\{\mathbf{f}_{k,l}, \mathbf{w}_{k,l}\}_{k=1}^K$ for each epoch $l$.

\STATE \textbf{Initialization: }model probabilities $\{\rho_{k,0}^{(i)}\}_{k=1}^K$, states $\{\hat{\mathbf{x}}_{k,0}^{(i)}\}_{k=1}^K$ and covariances $\{\hat{\mathbf{M}}_{0}^{(i)}\}_{k=1}^K$ for each user.

\FOR{each tracking epoch $l = 1,2,\dots$}
    \FOR{$k = 1$ to $K$}
        \STATE Predict state $\bar{\mathbf{x}}_{l|l-1}$ through IMM filter by employing (\ref{eqn:state_predict}) and design beamforming vectors $\mathbf{f}_{k,l},\mathbf{w}_{k,l}$ based on (\ref{eqn:tx_bf})-(\ref{eqn:rx_bf})
        \STATE Extract  $\mathcal{Z}_{k,l}$ via CFAR detection per (\ref{eqn:mea_vector})
        \STATE Compute association probabilities $\beta_q$ for each measurement following (\ref{eqn:asso_prob})
        \STATE Update  states $\hat{\mathbf{x}}_{k,l}^{(i)}$ and covariances $\hat{\mathbf{M}}_{l}^{(i)}$, and model probabilities $\{\rho_{k,l}^{(i)}\}_{k=1}^K$ $\forall i = 1,2$ according to (\ref{eqn:model_prob_update})
    \ENDFOR
\ENDFOR
\end{algorithmic}
\end{algorithm}

\begin{figure}[t]
\centerline{\includegraphics[width=6.5cm]{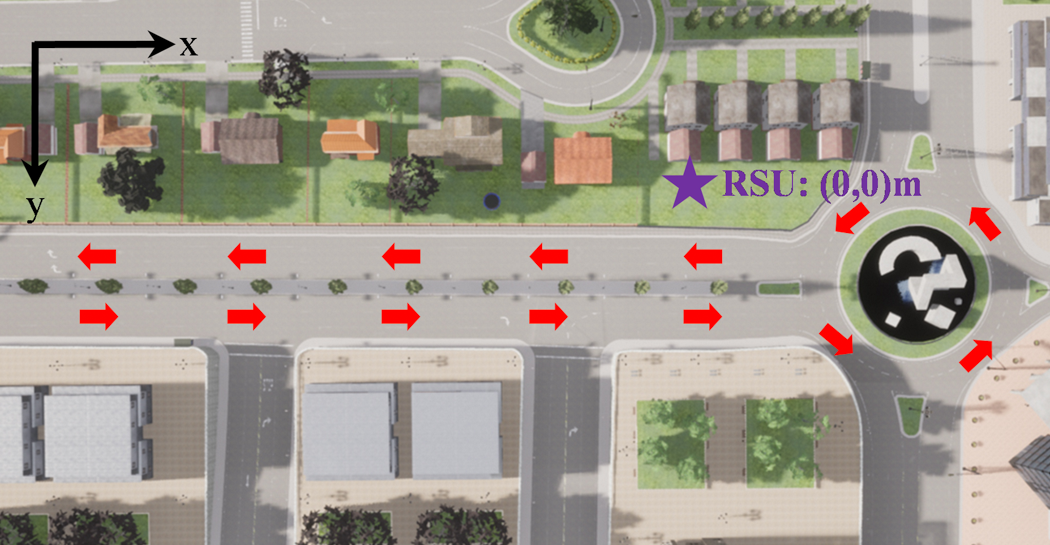}}
\caption{Generated scenario via CARLA for user association testing.}
\label{fig_simu}
\vspace{-6mm}
\end{figure}

\begin{figure}[t]
    \centering
    \subfloat[distance]{\includegraphics[width=4.3cm]{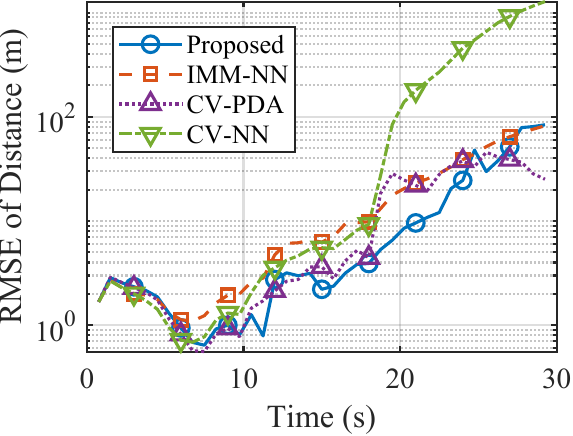}}
    \label{fig_tracking_dist}
    \hfil
    \subfloat[angle]{\includegraphics[width=4.3cm]{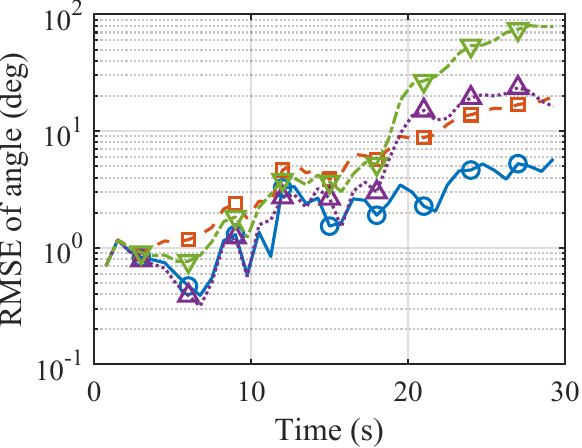}}
    \label{fig_tracking_ang}
    \caption{ Tracking error comparisons among different schemes.}
    \label{fig_tracking}
    \vspace{-5mm}
\end{figure}
\vspace{-1mm}
\section{Simulations}
\vspace{-1mm}
In this section, we provide numerical results based on Monte Carlo simulations to verify the effectiveness of the proposed framework. The dual-functional RSU is equipped with $N_t = N_r = 128$ antennas, operating at $f_c=30$ GHz with a bandwidth of $B=500$ MHz. The transmit power and noise power are set to $p_l=30$ dBm and $\sigma_n^2=\sigma_r^2=-57$ dBm respectively, similar to the configuration in \cite{meng2023vehicular}. The coherent processing time is set to $T=20$ ms, with matched filtering gain being $G=70$ dB. The path loss follows the free-space model with exponential factor as $1$. 

We consider three users with three clutter vehicles in the vicinity of each user. The tracking and association interval is set to $0.75$ s for each execution. For the IMM filter, we set the process noise standard deviations as $\sigma_{a_x}=\sigma_{a_y}=2$ m/s$^2$ for linear accelerations and $\sigma_{a_\omega}=3$ rad/s$^2$ for angular acceleration, while the measurement noise variances are set as $\sigma_\theta=0.005$ rad, $\sigma_r=0.02$ m, and $\sigma_v=0.01$ m/s. The vehicle trajectories are generated using CARLA simulator~\cite{Dosovitskiy17} to obtain realistic vehicle behaviors in complex road geometries, as shown in Fig. \ref{fig_simu}.

To evaluate the effectiveness of our proposed framework, we conduct ablation studies using four different schemes:
\begin{itemize}
    \item \textbf{Proposed}: Our proposed framework incorporating both IMM filtering and probabilistic data association.
    \item \textbf{IMM-NN}: IMM filtering with nearest neighbor association based on Euclidean distance, which helps validate the benefits of probabilistic association.
    \item \textbf{CV-PDA}: Single CV model with probabilistic data association, demonstrating the necessity of IMM filtering.
    \item \textbf{CV-NN}: Single CV model with nearest neighbor association, representing the oversimplified schemes.
\end{itemize}

As shown in Fig. \ref{fig_tracking}(a) and (b), our proposed method achieves superior tracking accuracy in both distance and angle prediction. The performance gain of IMM-NN over CV-NN validates the effectiveness of the IMM filter in handling complex vehicle maneuvers, while the improvement of CV-PDA over CV-NN demonstrates the advantage of probabilistic data association in cluttered environments. By incorporating both components, our IMM-PDA framework consistently outperforms all baselines, confirming the necessity of jointly addressing model mismatch and measurement uncertainty.

\begin{figure}[tp]
\centerline{\includegraphics[width=5.5cm]{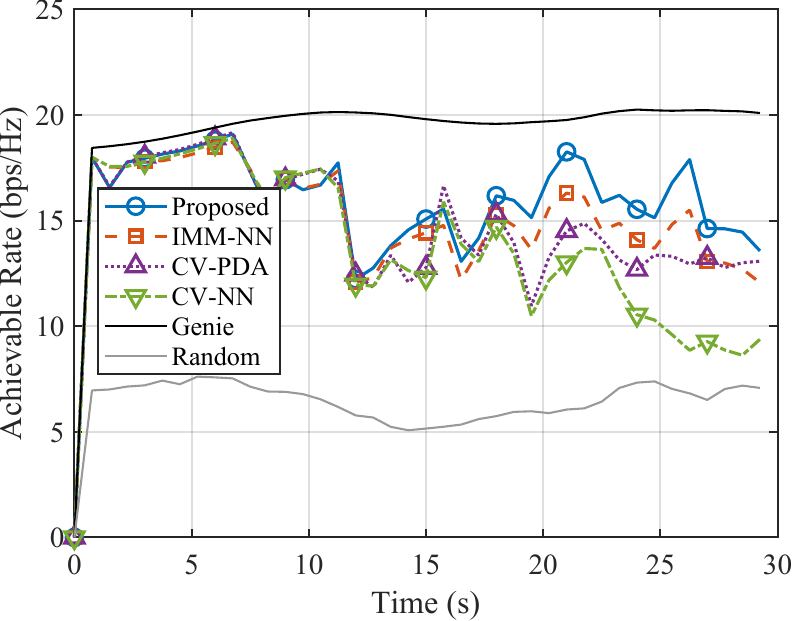}}

\caption{ Achievable rate comparisons among different schemes.}
\label{fig_rate}
\vspace{-4mm}
\end{figure}

Fig. \ref{fig_rate} demonstrates the achievable rate performance across different schemes. Two additional benchmarks are included: a Genie scheme that uses perfect state information for beamforming, serving as an upper bound, and a Random scheme with randomly selected beamforming vectors as a lower bound. Our proposed IMM-PDA framework achieves an average rate of $15.75$ bps/Hz, approaching $81.9\%$ of the genie benchmark ($19.22$ bps/Hz) while significantly outperforming the random scheme ($6.37$ bps/Hz). This performance advantage stems from more precise beam alignment enabled by enhanced tracking accuracy and reliable T2U association. Notably, during vehicle turns (around $t=10$s), all tracking-based schemes experience temporary performance degradation, but our IMM-PDA framework demonstrates faster recovery and better robustness than baselines.

\begin{figure}[tp]
\centerline{\includegraphics[width=5.5cm]{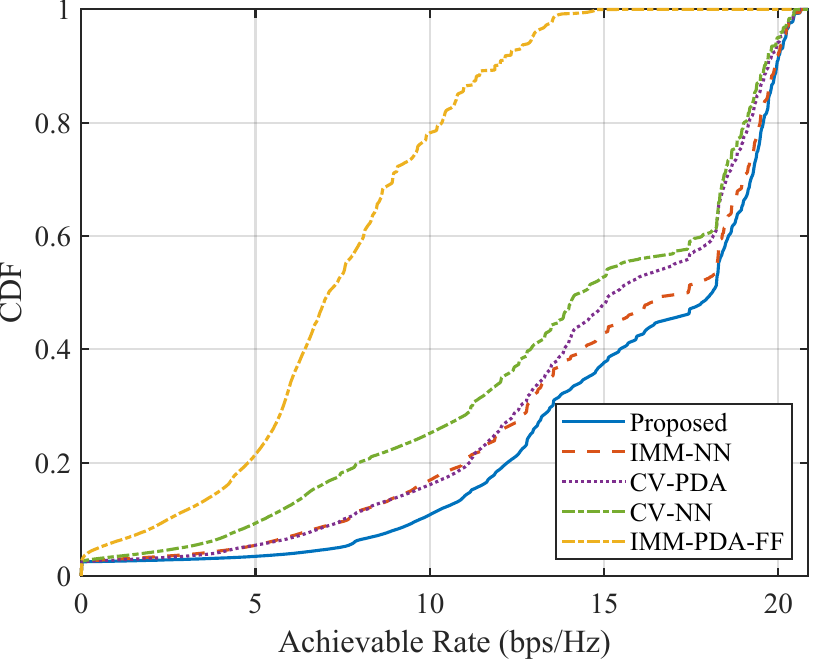}}
\caption{ CDF of achievable rates for different schemes.}
\label{fig_cdf}
\vspace{-6mm}
\end{figure}

In Fig. \ref{fig_cdf}, we compare the empirical cumulative distribution function (CDF) of achievable rates to evaluate communication reliability. We additionally include a variant of our framework that employs the far-field beamforming (IMM-PDA-FF) to demonstrate the importance of unified spherical wave modeling. The proposed framework consistently outperforms all schemes, while IMM-PDA-FF shows significantly degraded performance despite using the same algorithm, achieving an outage probability of $15.14\%$ at $4$ bps/Hz compared to $3.16\%$ for our unified model. This substantial gap (nearly $5$ times higher outage probability) validates the necessity of considering hybrid field effects. The remaining baselines IMM-NN, CV-PDA, and CV-NN achieve outage probabilities of $4.44\%$, $4.20\%$, and $6.66\%$ respectively, all outperforming the far-field variant but falling short of our proposed framework.
\vspace{-1mm}
\section{Conclusion}
\vspace{-1mm}
In this paper, we have presented a robust user association scheme for V2I systems. Our proposed approach utilizes ISAC signaling without the need for additional radar units. Furthermore, it combines IMM-based tracking with probabilistic data association to effectively handle complex vehicle maneuvers and reduce clutter interference. The framework demonstrates significant performance improvements, achieving a $5\%$ increase in average achievable rate and a $28.8\%$ reduction in outage probability at $4$ bps/Hz compared to the best benchmark scheme. These advancements highlight the effectiveness of our unified spherical wave modeling and predictive T2U association framework in hybrid-field scenarios. Future work will focus on expanding this framework into a comprehensive beam management pipeline that integrates multi-modal sensing for initial access and blockage prediction. Additionally, special attention will be paid to mitigating beam squint effects in hybrid-field scenarios to further enhance system performance.
\vspace{-4mm}

\bibliographystyle{IEEEtran}  
\bibliography{references}  

\end{document}